\documentclass[12 pt]{extarticle}
\usepackage{amssymb,amsmath,amsfonts,enumerate}
\usepackage{graphicx}
\usepackage{subcaption}

\usepackage{boxedminipage}
\usepackage{bm}
\usepackage{float}

\def\clock{{\count0=\time
           \divide\count0 60
           \ifnum\count0<10 0\fi\the\count0
           \multiply\count0 -60 \advance\count0 \time
           :\ifnum\count0<10 0\fi \the\count0
         }}
\newcommand{\timestamp}{{\small\vbox{\hbox{\tt\jobname.tex}
\hbox{\the\day/\the\month/\the\year, \clock}}}}

\newcommand{\beq}{\begin{equation}}
\newcommand{\eeq}{\end{equation}}
\newcommand{\ben}{\begin{displaymath}}
\newcommand{\een}{\end{displaymath}}
\newcommand{\beqa}{\begin{eqnarray}}
\newcommand{\eeqa}{\end{eqnarray}}
\newcommand{\bea}{\begin{eqnarray}}
\newcommand{\eea}{\end{eqnarray}}
\newcommand{\bean}{\begin{eqnarray*}}
\newcommand{\eean}{\end{eqnarray*}}
\newcommand{\ba}{\begin{array}}
\newcommand{\ea}{\end{array}}
\newcommand{\bi}{\begin{itemize}}
\newcommand{\ei}{\end{itemize}}

\numberwithin{equation}{section}

\begin{document}

\begin{titlepage}
\begin{flushright}
\end{flushright}
\vskip 2.cm
\begin{center}
{\bf\LARGE{Solar System, Astrophysics, and Cosmology from the Derivative Expansion}} 
\vskip 1.5cm

{\bf Nidal Haddad$^1$ and Fateen Haddad$^2$
}
\vskip 0.5cm
\medskip
\textit{ $^1$Department of Physics, Bethlehem University}\\
\textit{P.O.Box 9, Bethlehem, Palestine}\\

\vskip .2 in
\texttt{haddad.nidal02@gmail.com, f.a.assaleh@gmail.com}

\end{center}

\vskip 0.3in

\baselineskip 16pt
\date{}

\begin{center} {\bf Abstract} \end{center} 

\vskip 0.2cm 

\noindent 
In this paper we show how the solar system, the galactic, and the cosmological scales, are accommodated in a single framework, namely, in the derivative expansion framework. We construct a locally inertial static metric, based on the Einstein equations and on the derivative expansion method, which describes a Schwarzschild black hole immersed in dark matter and dark energy. The leading order metric in the expansion corresponds to the solar system, the first order metric to the galaxy, and the second order metric to cosmology. It is shown how this metric captures the main observations at each scale: in the solar system it trivially gives the Keplerian physics, in the galaxy it gives the flat part of the rotation curve and the Baryonic Tully-Fisher relation, and in the cosmological scale it gives the cosmological redshift, the accelerating expansion, and it coincides with the Robertson-Walker spacetime in the appropriate limit and approximation.  
\end{titlepage} \vfill\eject

\setcounter{equation}{0}

\pagestyle{empty}
\small
\normalsize
\pagestyle{plain}
\setcounter{page}{1}

\newpage
\section{Introduction}

The fact that a general coordinate transformation does not change the physics in General Relativity (the principle of general covariance) does not mean that all coordinate systems are of the same importance and relevance to our observations of the universe. After all, we are locally inertial observers of the universe (the principle of equivalence) and therefore using a locally inertial coordinate system would be the most appropriate for our observations. The Schwarzschild metric in the standard form is one fundamental solution of the Einstein equations that is written in locally inertial coordinates; the locally inertial observer is sitting far away from the spherical gravitational source (e.g., a star, a black hole, etc.) where the metric is approximately Minkowski. The Schwarzschild metric is the basis for the classical tests of General Relativity, and it is especially successful when applied to planetary motions about the Sun. This metric describes, however, an isolated source (from the rest of the universe) as the asymptotes are flat. On the other hand, the Friedmann–Lemaître–Robertson–Walker metric (in short FLRW) is another fundamental solution of the Einstein equations which describes the large scale structure of the universe such as the cosmological redshift and the accelerating expansion, but it is written in global coordinates and not in locally inertial ones. The search for a  locally inertial metric that accommodates both the local observations (the physics around a nearby spherical source, given by the Schwarzschild metric) and the cosmological observations (given by the FLRW spacetime) is not new, but almost as old as General Relativity is (see for example \cite{McVittie:1933zz,Einstein:1945id,Cooperstock:1998ny,Mizony:2004sh,Faraoni:2007es,Mitra:2013rma} and the references therein). 

In this paper we introduce a new approach for solving this problem, based on the derivative expansion method applied to the Einstein equations. The idea is to take the Schwarzschild black hole as the seed metric that describes the local physics and build over it a more general metric describing a black hole immersed in dark matter and dark energy (being the main constituents of the universe)\footnote{In General Relativity it is common to approach problems concerning local physics by using the Fermi coordinates, where one builds the approximate local metric over the Minkowski spacetime; see \cite{Manasse:1963zz,Ni:1978zz,Marzlin:1994wc,Misner:1973prb} and see for example \cite{Pajer:2013ana,Dai:2015jaa,Cabass:2016cgp,Dai:2015rda} for modern applications and modifications in cosmology. In our problem we found it easier, more practical, and more intuitive to start immediately with the Schwarzschild black hole and build the approximate local metric over it. }. To observe the effects of dark matter and dark energy large distances are required from the point of view of the locally inertial observer, and similarly, significant changes in the cosmos due to dark matter and dark energy require large distances as well with respect to this observer. Thus, the contributions of dark matter and dark energy to our (locally inertial) metric seem to accept being expanded in derivatives; that is, the derivatives of those contributions to the metric seem to be very small compared to the local scales. We argue in this paper that the metric up to second order in derivatives is static and we show that: (1) The leading (zeroth) order metric in the expansion is the Schwarzschild metric corresponding to the solar system scale\footnote{More generally, it corresponds to any spherical static source as long as the asymptotes are Keplerian.}, (2) The first order metric corresponds to the galactic scale (the physics due to dark matter), as it gives the main features of the rotation curve of galaxies, like the flat part and the Baryonic Tully-Fisher relation \cite{Tully:1977fu,McGaugh:2000sr,McGaugh:2011ac}, and (3) The second order metric corresponds to the cosmological scale (the physics due to dark energy) and it is shown to coincide with the FLRW spacetime in the appropriate limit and approximation, and it gives rise to the cosmological redshift and the accelerating expansion.   

The intermediate scale (the galactic scale) was missed in the previous attempts at attacking this problem (in the references \cite{McVittie:1933zz,Einstein:1945id,Cooperstock:1998ny,Mizony:2004sh,Faraoni:2007es,Mitra:2013rma} and subsequent works), whereas in this paper it is shown to be an important matching region between the solar system and cosmological scale. In this sense, one of the main results of this paper is to put the three fundamental scales (solar system, galactic, and cosmological) consistently in a single framework; in the derivative expansion framework.

An interesting result of this analysis is to show, qualitatively and quantitatively, from the locally inertial metric found in this paper, how the anisotropies in space die out as we move up from the galactic scale to the cosmological one, where the metric goes to the FLRW spacetime in the cosmological limit.  Relatedly, the energy-momentum tensor of dark matter and dark energy sourcing this spacetime is found, with a nontrivial and interesting $r$-dependence, that is,  it encodes information about dark matter and dark energy at all distances. The energy-momentum tensor is that of an anisotropic fluid and it is shown to reduce to an isotropic fluid in the cosmological limit.

The paper is organised as follows. In Sec.[\ref{DerEx}] we introduce the derivative expansion method and apply it to build our locally inertial metric up to second order in derivatives. The leading order terms of the metric are shown to correspond to the solar system, the first order terms to astrophysics of galaxies, and the second order terms to cosmology. In Sec.[\ref{RW}] we show how the derivative expansion method is applied to the FLRW metric and how it reduces to a static local form if we change coordinates and perform the expansion up to second order in derivatives. In Sec.[\ref{redsh}] we show how the cosmological redshift is calculated from the locally inertial static metric. Sec.[\ref{Diss}] is for discussion and it is followed by some appendices.

\section {The Derivative Expansion in the Local Frame}
\label{DerEx}
The Schwarzschild solution of the Einstein equations
\beq\label{}
ds^2=-\left(1-\frac{2GM}{r}\right)dt^2+\left(1-\frac{2GM}{r}\right)^{-1}dr^2+r^2\left(d\theta^2+\sin^2\theta d\phi^2\right)
\eeq  
 is an example of a locally inertial metric describing local physics. The black hole, the star, or in general, the spherical source, is at the centre of the coordinate system and the locally inertial observer is located far away, at $r>>2GM$, where the metric is almost Minkowski. Observers on Earth, for example, observing planets bound to the Sun use this metric. Nevertheless, the Schwarzschild metric is asymptotically flat, which means among other things, that it describes the gravitational field of a source isolated from the rest of the universe -- i.e., isolated from the environment. But the solar system is immersed 
in the galaxy, the galaxy is immersed in the cluster, and the cluster in the supercluster, and each larger structure certainly has an influence (however small) on the smaller structure. There is no doubt that these influences will change the asymptotic behaviour of the Schwarzschild metric of the local observer, and "they will be more detectable if the local region of the observer is made large enough". 

In this paper our focus will be on the influence of dark matter and dark energy on the local metric. That is, we will construct a solution to the Einstein equations which describes a black hole immersed in dark matter and dark energy. As is well observed, considerable changes in the local metric due to dark matter and dark energy require very large distances relative to the local observer.\footnote{The units of $\ kpc$ and $\ Mpc$ used in galactic astrophysics and cosmology imply how large distances are required to see changes.} This leads us to expect that the derivative expansion method is supposed to work and be successful in this problem.

Dark matter is known to be spherically symmetric around galaxies (such distributions are called dark halos). Dark energy is known to be uniformly distributed in the universe, and so it is supposed to be spherically symmetric around local observers. This leads us to look for a solution with a spherical symmetry. 
Furthermore, for a local observer, changes in galaxies and in the cosmos demand very long times and so we are guided to look for a static metric. In fact, there is another reason to look for a static metric and it is a result that we will discuss in  Sec.[\ref{RW}]: The FLRW metric if written in the local coordinates it will be static up to order $H_0^2$, where $H_0$ is the Hubble's constant. Thus, if the metric of the largest scale (describing an expanding universe) is static in the local coordinates, then the metric at smaller scales has a greater reason to be so. The staticity assumption is also supported by some works like \cite{Einstein:1945id,Mizony:2004sh}.

The general static and spherical metric can be put in the form (see for instance \cite{Weinberg:1972kfs,Wald:1984rg}),
\beq\label{metric1}
ds^2=-f(r)dt^2+h(r)dr^2+r^2\left(d\theta^2+\sin^2\theta d\phi^2\right)
\eeq  
where we find it suitable for our purposes to write the functions $f(r)$ and $h(r)$ as
\beq\label{fh}
f(r)=1-\frac{2GM_b}{r}+V(r) \qquad \qquad h(r)=\left(1-\frac{2GM_b}{r}+U(r)\right)^{-1}
\eeq  
where $M_b$ is the baryonic mass of the local source, while $V(r)$ and $U(r)$ are two unknown functions. The two functions $V(r)$ and $U(r)$ clearly encode information about the background in which the local source is immersed; they are the contributions of dark matter and dark energy to the locally inertial metric. 
\subsection {Introducing the Derivative Expansion}
\label{introDE}
Let us Taylor expand the contributions of dark matter and dark energy -- the functions $V(r)$ and $U(r)$ -- around some radius. Since we want to impose horizon regularity on our solution we find it suitable to expand about the black hole radius $r=r_0$: \footnote{Later on we will show that expanding about the horizon's radius makes it easier to impose horizon regularity.}
\begin{align}\label{V}
V(r)&=V_0+V_1(r-r_0)+\frac{1}{2}V_2(r-r_0)^2+...\\\label{U}
U(r)&=U_0+U_1(r-r_0)+\frac{1}{2}U_2(r-r_0)^2+...
\end{align}
where $V_0=V(r_0)$ , $V_1=dV(r_0)/dr$, $V_2=d^2V(r_0)/dr^2$, etc., and the same for $U(r)$. Because the derivatives are very small the \textbf{generic} hierarchy in the Taylor terms\footnote{We have said that this is a "generic"  hierarchy because there are special cases where for example $V_0=0$, or $V_1=0$, or some other higher derivatives vanish at $r_0$. This of course does not ruin the hierarchy in the Taylor series, because then the relevant term simply does not exist.} 
\beq\label{}
V_0>>V_1(r-r_0)>>\frac{1}{2}V_2(r-r_0)^2>>...
\eeq  
\beq\label{}
U_0>>U_1(r-r_0)>>\frac{1}{2}U_2(r-r_0)^2>>...
\eeq  
continues to hold even for large $r$'s; and hence one can consider only the first few terms of the Taylor series (depending on the desired precision) even for large radial coordinate $r$. This is one of the important benefits of the derivative expansion that we will rely on in this work. 

An additional important and useful feature of the derivative expansion is that if you increase $r$ so much that the third term in the Taylor series, $\frac{1}{2}V_2(r-r_0)^2$, becomes larger than the first two terms ($V_0$ and $V_1(r-r_0)$), the higher order terms may still be negligible and the Taylor series with only the first three terms may still be an excellent approximation; this will be the case if the derivatives  $V_3,V_4,...$ are sufficiently small. This feature will play an essential role in our analysis, when we move from the solar system scale to the galactic scale and from the galactic scale to the cosmological scale.

One of the main goals of this paper is to fix the unknown constants $V_0,U_0,V_1,U_1,V_2,U_2$ from observational data and from physical principles.
\subsection{Regularity of the Horizon}
It is important to make sure that the black hole has a regular horizon, so as to prevent any absurd behaviour in the black hole vicinity. In order to have a regular horizon the functions $f(r)$ and $h(r)$ must satisfy (see Appendix \ref{app:Edd} for details),
\beq\label{}
f(r)=h(r)^{-1} \quad \textrm{as} \quad r\rightarrow r_0  
\eeq 
which is satisfied only if
 \beq\label{}
U_0=V_0
\eeq 
The position of the horizon is obtained from the equation $h(r)^{-1}=0$, and it is easily found to be, 
 \beq\label{}
r_0=\frac{2GM_b}{1+V_0}
\eeq 
As the presence of dark matter and dark energy is supposed to increase the mass inside the black hole, and its radius too, we conclude that $V_0<0$. 

\subsection{Zeroth Order in Derivatives (Solar System)}
At the leading order, with no derivatives, the functions $f(r)$ and $h(r)$ read, after imposing horizon regularity, 
\beq\label{}
f(r)=1-\frac{2GM_b}{r}+V_0 \qquad \qquad h(r)=\left(1-\frac{2GM_b}{r}+V_0\right)^{-1}
\eeq  
The constant $V_0$ is much smaller than the Newtonian term $2GM_b/r$ in the solar system since the effects of dark matter and dark energy are subdominant there compared to Newtonian physics.\footnote{The constant $V_0$ is a constant shift in the potential energy, however, in contrast to Newtonian physics and special relativity, this constant can not be removed in general relativity; because it induces some curvature.}

Also in astrophysics and cosmology, that is, for very large $r$'s, we can  neglect $V_0$ in front of the growing first order terms, $V_1(r-r_0)$ and $U_1(r-r_0)$, and second order terms, $\frac{1}{2}V_2(r-r_0)^2$ and $\frac{1}{2}U_2(r-r_0)^2$. Therefore, for all practical uses, it is justified to take,   
\beq\label{}
U_0=V_0=0
\eeq 
Hence, we see that at zeroth order in derivatives the Schwarzschild metric dominates, corresponding to the solar system scale, or in general, to the local source scale.

\subsection{First Order in Derivatives (Astrophysics)}
The first order calculation was done in reference \cite{Haddad:2020koo}, and it was shown that it corresponds to the galactic scale. In what follows we will briefly summarise the general lines and results of \cite{Haddad:2020koo}, and we will also sharpens some points along the way. The metric up to first order in derivatives reads,
\beq\label{}
ds^2=-\left(1-\frac{2GM_b}{r}+V_1(r-2GM_b)\right)dt^2+\frac{dr^2}{1-\frac{2GM_b}{r}+U_1(r-2GM_b)}+r^2d\Omega^2
\eeq 
where, as a simplified model for galaxies, we are assuming that all the baryonic mass of the galaxy, $M_b$, lies inside a central black hole, when this black hole in turn is immersed in a very large dark matter halo (the reservoir).  

Assuming that stars take circular orbits about the centre of the galaxy the rotation curve can be shown to take the simple form (see Appendix \ref{app:geod}),
 \beq\label{}
v^2=\frac{GM_b}{r}+\frac{1}{2}V_1 r
\eeq 
where the first term is the well-known baryonic (or Keplerian) term and where the second is the new term arising from dark matter. The constant $V_1$ must clearly be positive ($V_1>0$) since otherwise the rotation curve will be a monotonically decreasing function of $r$, at a rate that is even faster than the rate of $GM_b/r$ alone, and so a constant (or almost constant) rotation curve can not be obtained. Furthermore, if we use the basic circular motion relation $F_r=-mv^2/r$ we get that $F_r=-GM_bm/r^2-mV_1/2$ and so we see that dark matter induces a constant acceleration toward the centre of magnitude $V_1/2$ in the Newtonian limit. We will call the magnitude of this constant acceleration $a_0$,
 \beq\label{}
a_0\equiv V_1/2
\eeq 
and from now on we find it more convenient to work with $a_0$ instead of $V_1$ since it has a more direct physical meaning. Then the rotation curve is written as,
 \beq\label{rotcurve}
v^2=\frac{GM_b}{r}+a_0 r
\eeq 
  
\begin{figure}[]
\begin{center}
\includegraphics[bb=100 250 500 620 ,scale=0.5]{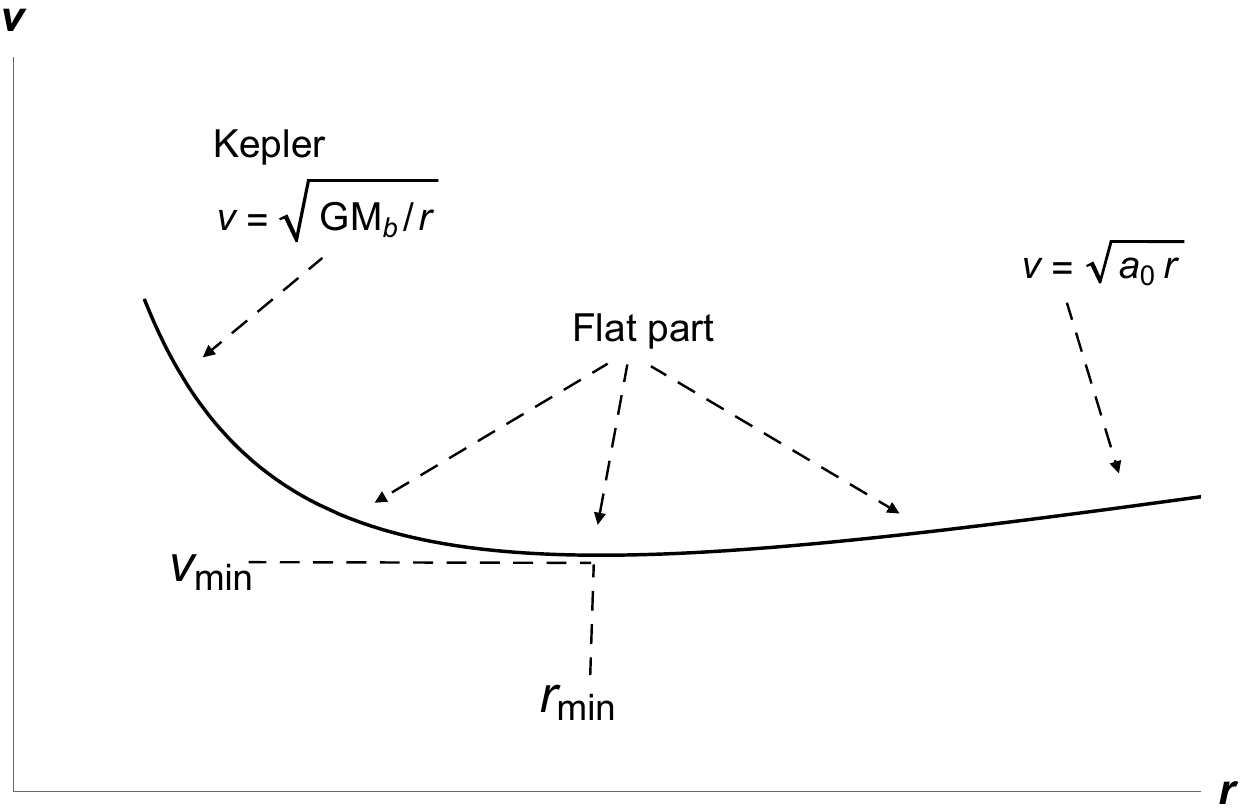}
\end{center}
\caption{\small This is a plot of the rotation curve $v=\sqrt{\frac{GM_b}{r}+a_0r}$. Three parts are clear: the Keplerian part, the flat part around the minimum, and the increasing $\sqrt{a_0r}$ part. In galaxies there are no big amounts of baryonic matter outside the flat part and that is why we are not disturbed by the third part. The key is that the constant rotation curve of galaxies is a result of a very flat minimum. }\label{fig:vr}
\end{figure}

Since in our derivative expansion the derivative $V_1$ (or equivalently $a_0$) is assumed small we see that for small $r$'s the Keplerian term $GM_b/r$ is much larger than the linear term $a_0r$ and thus we obtain the familiar Keplerian rotation curve $v=\sqrt{GM_b/r}$. As we increase $r$ the Keplerian term $GM_b/r$ continues in decreasing while the linear term $a_0r$ keeps increasing until a minimum velocity is reached after which a monotonic increase in the velocity is seen (see $Fig$.\ref{fig:vr}). For large enough $r$'s the linear term dominates and the rotation curve becomes $v=\sqrt{a_0r}$. In fact, the flat rotation curve will appear naturally around the minimum velocity because firstly the slope is zero there and secondly because the curvature of the minimum is very small as will be shown as we proceed. In other words, the constant rotation curve of galaxies is a result of a very flat minimum.

The location of the minimum velocity (the location of the centre of the flat part of the rotation curve) is obtained from the equation $dv/dr=0$, which upon solving gives,
 \beq\label{}
r_{min}=\sqrt{\frac{GM_b}{a_0}} \quad \quad \textrm{(the location of the centre of the flat part)}
\eeq 
Note that the larger the mass of the galaxy the more distant the flat part is, and note that a smaller $a_0$ makes the flat part further away. The minimum velocity  (the constant velocity of the rotation curve) can then be easily computed,
\beq\label{}
v_{min}=\left(4GM_ba_0\right)^{1/4} \quad \quad \textrm{(the flat velocity)}
\eeq 
This last result can be reversed and rewritten as,
\beq\label{BTFR}
M_b=\frac{v_{min}^4}{4Ga_0} \quad \quad \textrm{(Baryonic Tully-Fisher relation)}
\eeq 
which can be identified with the Baryonic Tully-Fisher relation ($M_b\propto v^4$, see \cite{Tully:1977fu,McGaugh:2000sr,McGaugh:2011ac} ), if and only if $a_0$ is a universal constant (that does not depend on $M_b$). Therefore, we see that the empirical Baryonic Tully-Fisher relation requires that,
\beq\label{}
a_0=(\textrm{universal constant})
\eeq 
 This last statement might not sound strange in our picture because  $a_0$ (or equivalently $V_1$) is supposed to depend on the background fields (in this case on the reservoir of dark matter) and not on the local small system (the baryonic disk of the galaxy); if so then the Baryonic Tully-Fisher relation can be considered as a prediction of our analysis.

As for the width of the flat part of the rotation curve one can calculate the curvature of the curve about the minimum and find\footnote{We have divided the second derivative by $v_{min}$ to obtain the correct units of curvature, $1/\textrm{length}^2$.},
  \beq\label{}
\frac{1}{v_{min}}\left(\frac{d^2v}{dr^2}\right)_{r_{min}}=\frac{a_0}{4 GM_b}=\frac{1}{4r_{min}^2}
\eeq 
which is clearly small as it is proportional to $a_0$. In more details, if we expand about the minimum, $r=r_{min}+\Delta r$,  we get that the velocity is 
 \beq\label{}
v=v_{min}\left(1+\frac{1}{8}\left(\frac{\Delta r}{r_{min}}\right)^2\right)+...
\eeq 
where the factor of $1/8$, the squared ratio $\left(\frac{\Delta r}{r_{min}}\right)^2$, and the large value of $r_{min}$, all together, tell that the flat part is noticeably wide. Quantitatively, for a typical spiral galaxy the order of magnitude of the flat velocity is $v_{min}\sim 10^2\ km/s$ and the order of magnitude of the location of the centre of the flat part is $r_{min}\sim 10 \ kpc$. So that, for a displacement $\Delta r\sim 1 \ kpc$ one has $(v-v_{min})\sim10^{-1}\ km/s$ which clearly indicates the existence of a flat minimum in galactic scales (a constant rotation curve).

As for the order of magnitude of the universal constant $a_0$ it was found to be 
\beq\label{}
 a_0\sim 10^{-11}\ m/s^2
 \eeq
 This order of magnitude of $a_0$ was found by using real data as follows. A typical spiral galaxy has the following approximate ranges for its parameters: the velocity in the flat part $v\in [200,250] \ km/s$, the baryonic mass $M_b\in [0.5\times 10^{11},2\times 10^{11}]M_{sun}$, and take a star that lies in the flat part in the range $r\in [30,40]\ kpc$  (see for example the review \cite{Sofue:2017}). With these ranges of parameters, by using $Eq$.(\ref{rotcurve}) to extract $a_0$, one gets that $a_0\sim 10^{-11}\ m/s^2$. We have chosen deliberately wide ranges for the parameters to show that the order of magnitude of $a_0$ is insensitive which gives further indication that it is a universal constant.

The smallness of $a_0$ is an indication of the correctness of the derivative expansion method, in the sense that this makes the linear term $a_0r$ (or equivalently $\frac{1}{2}V_1r$) change slowly along galactic distances (several kilo-parsecs) and also in the sense that several kilo-parsecs are required before this linear term becomes comparable with the baryonic term $GM_b/r$. 

\begin{figure}[H]
\centering

\begin{minipage}[b]{0.45\linewidth}
\includegraphics[bb=50 0 250 380 ,scale=0.45]{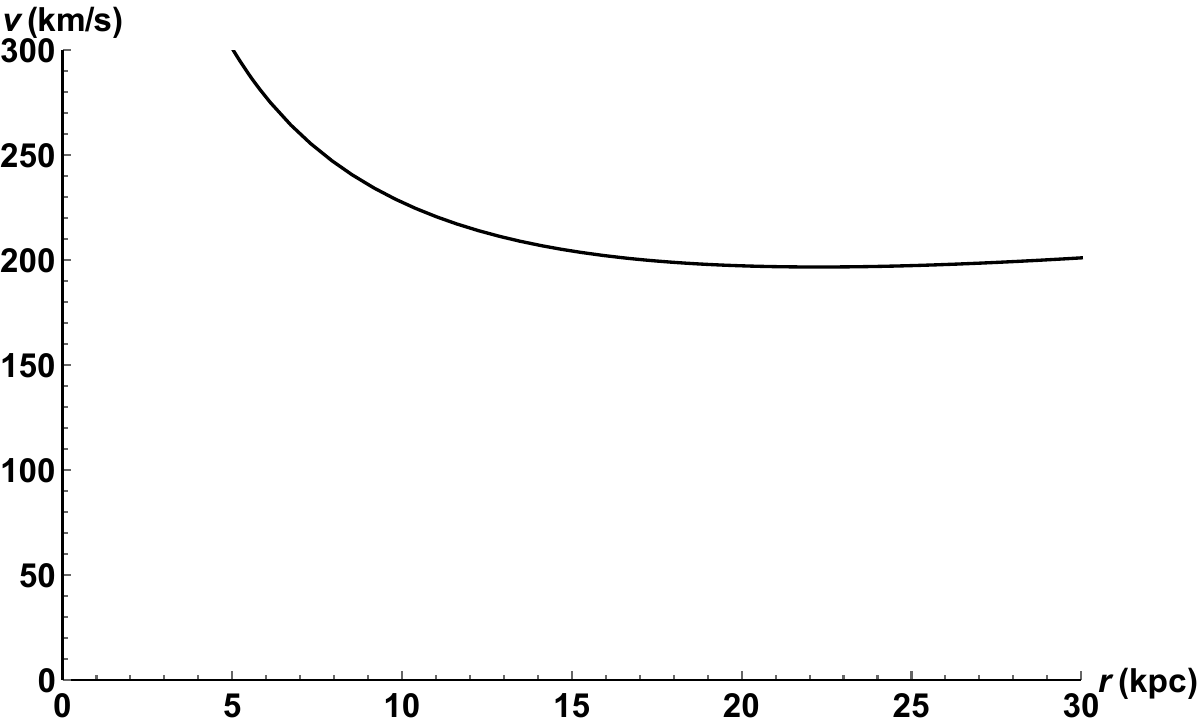}
\label{fig:minipage01}
\end{minipage}
\quad
\begin{minipage}[b]{0.45\linewidth}
\includegraphics[bb=0 0 250 380 ,scale=0.45]{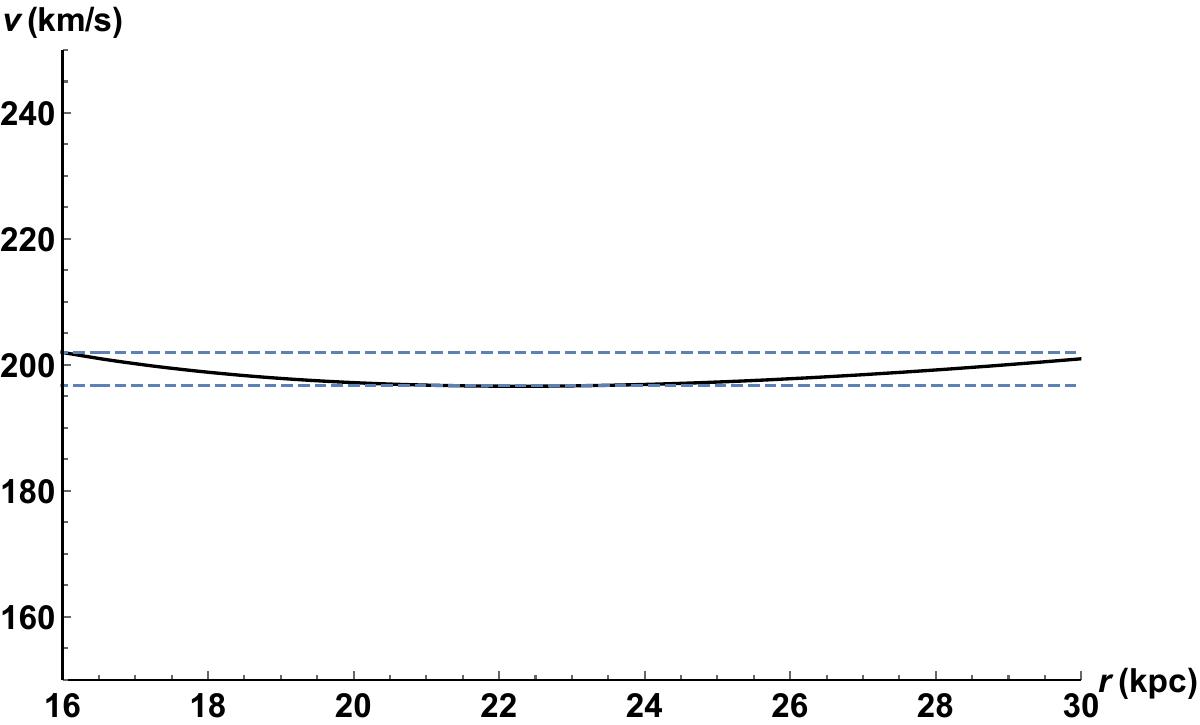}
\label{fig:minipage02}
\end{minipage}

\caption{\small The two plots are taken from reference \cite{Haddad:2020koo}. Left: This is a plot of the rotation curve of the Milky Way based on $Eq$.(\ref{rotcurve}), with parameters  $a_0=2.8\times 10^{-11}\  m/s^2$ and $M_b=10^{11}M_{sun}$. The flat part of the rotation curve is clear (concentrated at about $200\ km/s$) and also the Keplerian curve is clear just before the onset of the flat part.   Right: Here we zoom in on the flat part (solid line), extending from $16\ kpc$ to $30 \ kpc$ along which the velocity lies in the small range $[197\ km/s,202\ km/s]$ bounded by the two dashed lines. Here $r_{min}\approx 22\ kpc$ and $v_{min}\approx 197 \ km/s$.
 }

\label{fig:rotation curve}
\end{figure}

 As an illustration of all this a plot of the rotation curve of the Milky Way, based on $Eq$.(\ref{rotcurve}), is given in $Fig$.\ref{fig:rotation curve}, where we have made a rough estimation that\footnote{By using results from MOND, where there has been a lot of work and fittings to large numbers of galaxies, the proportionality factor in the Baryonic Tully-Fisher relation, $v^4=A_0M_b$, was found to be $A_0=Ga_{mond}$ with $a_{mond}=1.2\times 10^{-10}\ m/s^2$ with an uncertainty of a few tens of percents (see the reviews \cite{Milgrom:2014usa,Milgrom:2019cle} and the references therein). In our case -- see $Eq$.(\ref{BTFR}) -- we have obtained $A_0=4Ga_0$ and so we can calculate our $a_0$ to be $a_0=a_{mond}/4=3\times 10^{-11}\ m/s^2$ which is very close to our estimation.}
 \beq
 a_0=2.8\times 10^{-11} \ m/s^2 
 \eeq
 This rough estimation was made by using  reasonable approximate values for a single point on the rotation curve of the Milky Way: $r=30\ kpc$, $v=200\ km/s$, and $M_b=10^{11}M_{sun}$ (see references \cite{Sofue:2017,Sofue:2015}). By inserting these values in the rotation curve $Eq.$(\ref{rotcurve}) one obtains  $a_0=2.8\times 10^{-11} \ m/s^2$.

To fix the second constant $U_1$ we turn to the Einstein field equations $R_{\mu\nu}-\frac{1}{2}Rg_{\mu\nu}=8\pi G T_{\mu\nu}$. The time-time component gives the following energy density,
 \beq\label{}
\rho=\frac{-U_1\left(r-GM_b\right)}{4\pi G r^2}
\eeq 
which in the large radius (Newtonian) limit, $r>>2GM_b$, reduces to,
 \beq\label{density1}
\rho=\frac{-U_1}{4\pi G r}
\eeq 

On the other hand, in the Newtonian limit the Poisson equation  $\nabla^2\phi =4\pi G\rho$  governs the physics, where the gravitational potential for our static spacetime is given by $\phi\approx-(g_{00}+1)/2=GM_b/r+a_0r$. Thus, from the Poisson equation and for $r>0$ one obtains,
 \beq\label{density2}
\rho=\frac{a_0}{2\pi G r}
\eeq 
where since we are talking about large $r$'s the delta function is irrelevant.

Thus, by comparing $Eq$.(\ref{density1}) with $Eq$.(\ref{density2}) we identify\footnote{This is a boundary condition on the metric.},
\beq\label{}
U_1=-2a_0 
\eeq
As a summary, the two first-order constants in the derivative expansion are,
\beq\label{}
V_1=-U_1=2a_0 
\eeq
and the metric reads,
\beq\label{galaxy}
ds^2=-\left(1-\frac{2GM_b}{r}+2a_0(r-2GM_b)\right)dt^2+\frac{dr^2}{1-\frac{2GM_b}{r}-2a_0(r-2GM_b)}+r^2d\Omega^2
\eeq 

One can now easily calculate the energy-momentum tensor of dark matter (sourcing this spacetime) by simply plugging the above metric into the Einstein equations,  upon which one finds that,
\beq
T_\mu^\nu=\textrm{diagonal}(-\rho,P_r,P_\perp,P_\perp)
\eeq
with 
\beq
\rho=\frac{a_0}{2\pi Gr}-\frac{a_0M_b}{2\pi r^2} 
 \qquad \quad P_r=-\frac{a_0M_b}{2\pi r^2} \qquad\quad P_\perp=\frac{a_0M_b}{4\pi r^2}
\eeq
The energy-momentum tensor is that of an anisotropic fluid ($P_\perp\neq P_r$, see \cite{Bowers:1974tgi}) and was shown in \cite{Haddad:2020koo} to satisfy the four energy conditions (dominant, weak, null, and strong) outside the black hole horizon, making it clear that this is a physical matter field. Note also that the energy density of dark matter is much larger than the pressures for large radii, a result which is sometimes phrased in the literature by saying that "dark mater is pressure-less".  Close to the black hole, however, the density and pressures are of the same order of magnitude, signalling a region of relativistic dark matter.

\begin{figure}[H]
\centering
\begin{minipage}[b]{0.45\linewidth}
\includegraphics[bb=0 0 250 380 ,scale=0.43]{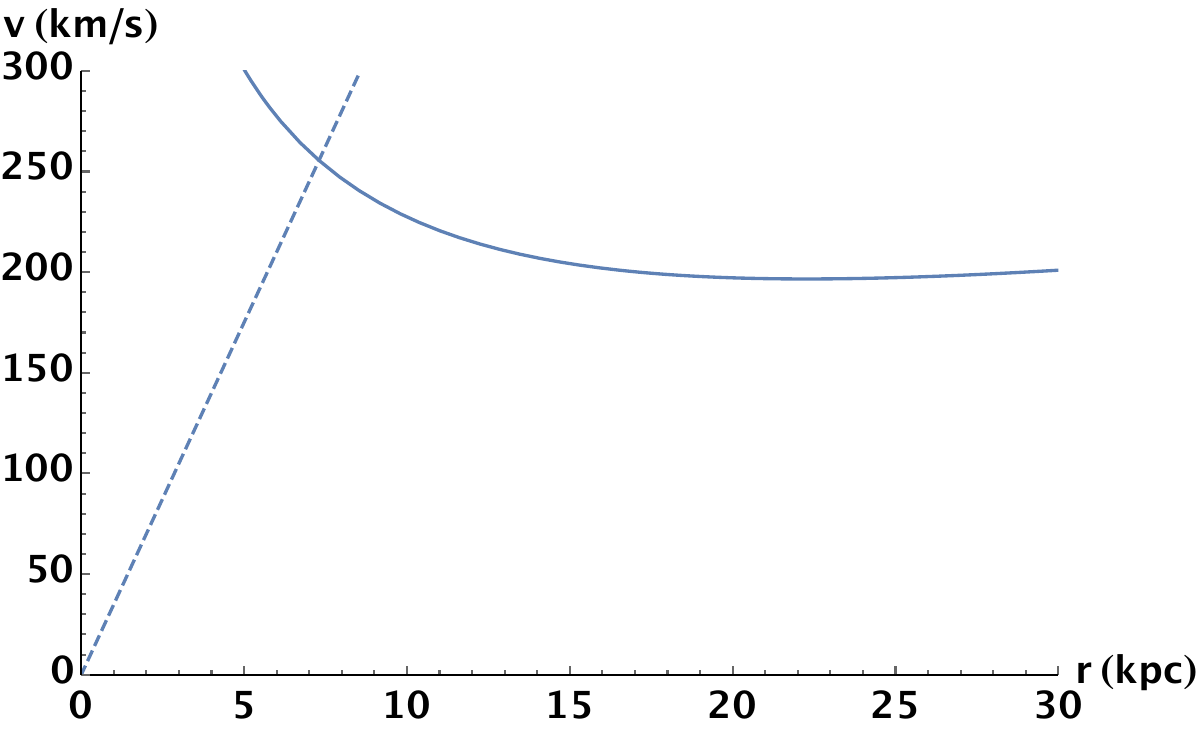}
\label{fig:minipage01}
\end{minipage}
\quad
\begin{minipage}[b]{0.45\linewidth}
\includegraphics[bb=0 0 250 380 ,scale=0.43]{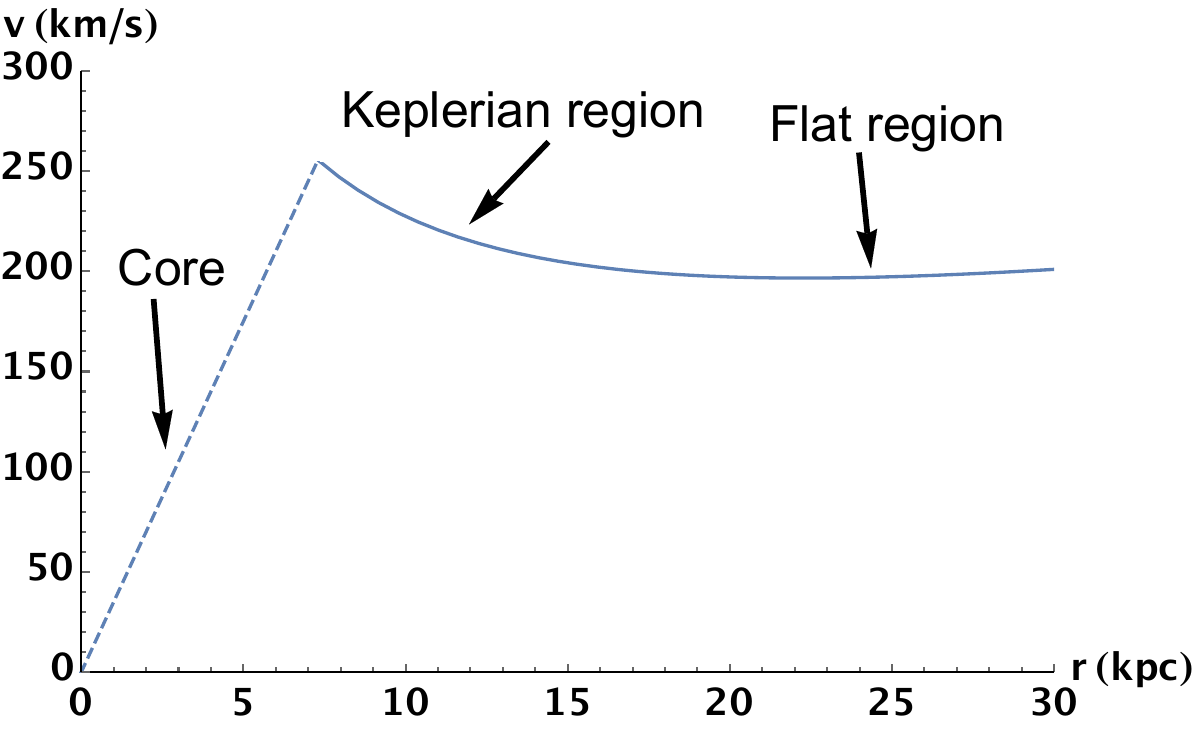}
\label{fig:minipage02}
\end{minipage}

\caption{\small Left: Our plot of the rotation curve (solid line) and the plot of the rotation curve of a rigid body (dashed line), which we took for illustration only, are put together. Right: Since our plot is valid outside the core and the linear curve (straight line) is the model we took inside, the two plots can be merged (in a rough way) together by simply taking off the additional parts after their intersection.
 }

\label{fig:FIG0}
\end{figure}

Finally, it is important to mention that our metric $Eq.$ (\ref{galaxy}) and the corresponding rotation curve $Eq.$(\ref{rotcurve}) describe a realistic model for galaxies only outside the cores inside which most of the baryonic matter is concentrated; for the milky way that would be about $r>5\ kpc$ (see reference \cite{Haddad:2020koo}). In our metric we are assuming -- as a simplified model -- that the baryonic matter of the core is concentrated inside the central black hole at $r=0$. However, if we want to be more precise inside the core we must take the real baryonic matter distribution there into account, that is, we must assume a certain mass density $\rho(r)$ for the baryonic matter inside the core and then solve the Einstein equations. As an illustration, if we assume a constant density of baryonic matter inside the core, $\rho = \textrm{constant}$, then, assuming non-relativistic physics, the velocity will increase linearly with distance like a rigid body, $v \propto r$. Then the combined rotation curve that contains the two regions, inside and outside the core, is given by $Fig.\ref{fig:FIG0}$,

This shows that the Keplerian rise (seen in our rotation curve) does not continue to the centre if real models in the core are assumed. The Keplerian rise in some galaxies (depending on the details) may be very small (barely seen) and in other galaxies may result in a clear peak in the rotation curve -- observed rotation curves exhibit these two possibilities.

\subsection{Second Order in Derivatives (Cosmology)}
In this section we will fix the second order constants  $V_2$ and $U_2$. To do so let us go to the largest scale, that is, to the limit "$r\rightarrow\infty$", where the second order terms $\frac{1}{2}V_2r^2$ and $\frac{1}{2}U_2r^2$ become dominant over the others in $Eq$.(\ref{V}) and Eq.(\ref{U}). The metric, Eq.(\ref{metric1}), in this limit reads,
\beq\label{metric-cosm}
ds^2=-\left(1+\frac{1}{2}V_2r^2\right)dt^2+\frac{dr^2}{1+\frac{1}{2}U_2r^2}+r^2\left(d\theta^2+\sin^2\theta d\phi^2\right)
\eeq  
In this limit our metric must coincide with the  Friedmann–Lemaître–Robertson–Walker spacetime (this is our boundary condition) since the latter is the spacetime which describes the largest scale; a homogeneous and isotropic universe. As we show in $Sec.$[\ref{RW}] if we take the FLRW metric, expand it in derivatives, and transform coordinates from global to local, then the metric up to second order in derivatives will read\footnote{This metric was already obtained in \cite{Mizony:2004sh} by using a different approximation method.},
 \beq\label{FLRWappro}
ds^2=-\left(1+q_0H_0^2r^2\right)dt^2+\frac{dr^2}{1-\Omega_0H_0^2r^2}+r^2\left(d\theta^2+\sin^2\theta d\phi^2\right)
\eeq
where $H_0$, $q_0$, and $\Omega_0$ are the well-known cosmological parameters at the present time: $H_0$ is the Hubble constant, $q_0$ is the deceleration parameter, and $\Omega_0$ is the dimensionless density parameter. Thus, the form of our limiting metric $Eq.$(\ref{metric-cosm}) successfully coincides with the "expanded" FLRW metric, $Eq.$(\ref{FLRWappro}), and hence we can identify our constants to be,    
 \beq\label{}
V_2=2q_0H_0^2 \qquad\textrm{and}\qquad U_2=-2\Omega_0H_0^2
\eeq
\subsection{Solar System, Astrophysics, and Cosmology in One Metric, and the Validity of the Derivative Expansion}
\label{Overview}
As an overview  we write below the local metric up to second order in derivatives, which accommodates the three distinct scales (solar system, astrophysics, and cosmology) in a single framework:
\begin{align}\label{Fullmetric}\nonumber
ds^2=&-\left[1-\frac{2GM_b}{r}+2a_0(r-2GM_b)+q_0H_0^2(r-2GM_b)^2\right]dt^2\\&
+\left[1-\frac{2GM_b}{r}-2a_0(r-2GM_b)-\Omega_0H_0^2(r-2GM_b)^2\right]^{-1}dr^2+r^2d\Omega^2
\end{align}
where the constants $M_b$, $a_0$, $H_0$, $q_0$, and $\Omega_0$ were defined before. 

\subsubsection*{Stars in Galaxies}
For stars in galaxies one can easily calculate the rotation curve up to second order (see Appendix \ref{app:geod}),
 \beq\label{} 
v_\phi^2=rf'(r)/2=GM_b/r+a_0r+q_0H_0^2r^2(1-2GM_b/r)
\eeq
In the far region of the baryonic black hole, $r>>2GM_b$, we can neglect the ratio $2GM_b/r$ in the third term and the rotation curve becomes 
 \beq\label{}
v_\phi^2=GM_b/r+a_0r+q_0H_0^2r^2
\eeq
The central force in the far region can be easily inferred from the circular motion equation $F_r=-mv^2/r$, which gives,
\beq\label{}
F_r=-GM_bm/r^2-ma_0-mq_0H_0^2r
\eeq
where note that the first two forces are attractive while the third force $-mq_0H_0^2r$ is repulsive because the accelerating expansion of the universe requires $q_0<0$.
\subsubsection*{Galaxies in Clusters}

For larger scales (clusters and superclusters) it is commonly assumed that there is no rotation. Galaxies do not rotate in clusters (in contrast to stars in galaxies) but they move radially with respect to the centre of the cluster with a velocity that is a combination of the Hubble flow and some peculiar velocity, $v_r=H_0r+v_{pec}$. Furthermore, in those large scales the baryonic mass contribution becomes less and less important. Now, the radial acceleration can be obtained from the geodesic equation (see Appendix \ref{geod1})
\beq\label{}
\frac{d^2r}{dt^2}+\left(\frac{h'}{2h}-\frac{f'}{f}\right)\left(\frac{dr}{dt}\right)^2-\frac{f^2J^2}{r^3h}+\frac{f'}{2h}=0
\eeq
Since there is no rotation we simply take $J=0$. Since the baryonic contribution is unimportant in those scales we can neglect all terms containing $M_b$. Since $dr/dt=v_r=H_0r+v_{pec}$ with $v_{pec}<<H_0r$, the above radial geodesic equation becomes up to second order,
\begin{align}\label{}
\frac{d^2r}{dt^2}&=-\frac{f'}{2h}=-a_0-\left(q_0H_0^2-2a_0^2\right)r
\end{align}
Finally, one can check that $a_0^2/H_0^2\sim 10^{-4}$ and so the $a_0^2$ term above can be neglected as well, and we are left with the desired result,
\begin{align}\label{acc}
\frac{d^2r}{dt^2}=-a_0-q_0H_0^2r
\end{align}
for the radial acceleration of galaxies in clusters; we see that at leading order galaxies freely fall toward the centre of the cluster with a constant acceleration $a_0\sim 10^{-11}\ m/s^2$ which tells, in particular, that the cluster is a gravitationally bound system. At next to leading order galaxies are participating in the expansion of the universe with a smaller acceleration $-q_0H_0^2r$. 
\subsubsection*{Clusters and Superclusters}
By the same logic one sees that $Eq.$(\ref{acc}) applies also to the motion of clusters in superclusters and to the motion of superclusters in the universe, with the difference that now the dark energy term $-q_0H_0^2r$ increases with each higher scale until it becomes comparable and finally larger than the dark matter term $-a_0$. The quantitive details follow in the coming section on the validity of the approximation. 

\subsubsection*{Accelerating Expansion}
As just said at the largest scale the repulsive force due to dark energy becomes dominant over the dark matter attractive force, and $Eq$.(\ref{acc}) becomes,
 \begin{align}\label{cos-acc}
\frac{d^2r}{dt^2}=-q_0H_0^2r
\end{align}
which is the manifestation of the accelerating expansion of the universe in the local metric.
In more details, if we  use the definition of the deceleration parameter $q_0=-\ddot R_0/R_0H_0^2$ then the last equation can be rewritten as,
\beq\label{}
\frac{d^2r}{dt^2}/r=\ddot R_0/R_0
\eeq
where remember that the "dot" indicates a derivative with respect to the cosmological time of the FLRW metric. For the local observer this is the statement that distant galaxies are co-moving with the expanding universe\footnote{By the assumption of homogeneity every observer would see the same physics and so we conclude that all galaxies are co-moving with the expanding universe, and not only the distant ones. For the local observer the expansion of near galaxies with the universe is masked by the local physics of dark matter.}.

\subsubsection*{Validity of the Approximation Method}
It is to be noted that the smallness of the first and second derivatives $V_1$, $U_1$, $V_2$ and $U_2$ justifies (and gives strength to) the use of the derivative expansion method. At distances of order $10\ kpc$ (typical size of galaxies), one can check that 
 \beq\label{}
1>>a_0r\sim 10^{-7}>>H_0^2r^2\sim 10^{-11},
\eeq
which is a good start for the derivative expansion method. This tells that it is justified indeed to reach galactic distances by the derivative expansion method. At distances of order $1\ Mpc$ (the typical size of a galaxy cluster) the linear dark matter term is still dominant over the Hubble's term,
 \beq\label{}
 a_0r\sim 10^{-5}>>H_0^2r^2\sim 10^{-7}, 
 \eeq
 which shows that the derivative expansion is still valid and which explains why galaxies in clusters are gravitationally bound systems; from $Eq$.(\ref{acc}) we see that dark matter induces an attractive central force $F_r=-ma_0$ that is much larger than the repulsive force induced by the Hubble's term, $F_r=-mq_0H_0^2r$. 
At distances of order $10\ Mpc$ (the typical size of a supercluster) the linear term $a_0r$ is larger than the Hubble's term $H_0^2r^2$ by one order of magnitude,
 \beq\label{}
a_0r\sim 10^{-4}>H_0^2r^2\sim 10^{-5}
\eeq
This again explains why superclusters are gravitationally bound systems as the attractive force due to dark matter is still larger than the repulsive force due to dark energy (see again $Eq$.(\ref{acc})).
At distances of order $100\ Mpc$ (the typical size of  a large supercluster and larger) the two terms become of the same order\footnote{
As notified in the beginning of the section ($Sec$.[\ref{introDE}]) having $a_0r\sim H_0^2r^2$ does not invalidate our approximate solution as long as the higher derivatives ($V_3,U_3,V_4,U_4,...$) are small enough; the Taylor series with the first three terms, $V(r)=V_0+V_1(r-r_0)+\frac{1}{2}V_2(r-r_0)^2$, would still be a good approximation.},
 \beq\label{}
a_0r\sim H_0^2r^2\sim 10^{-3}
\eeq
and gravitational systems are not necessarily bound anymore, because the attraction and repulsion are of same order of magnitude. At a distance of order $1000\ Mpc$ (scales much larger than superclusters) the Hubble's term finally dominates,
 \beq\label{}
 H_0^2r^2\sim 10^{-1}>a_0r\sim 10^{-2}
 \eeq
 This shows that with the derivative expansion method we can indeed reach large cosmological distances, up to order $1000\ Mpc$.
At this stage the expansion of the universe appears in a clear way as there are no bound systems anymore, but everything is flowing with the expansion of universe; because now -- see again $Eq$.(\ref{acc}) -- the repulsive force due to dark energy, $F_r=-q_0H_0^2r$, becomes stronger than the attractive force due to dark matter, $F_r=-a_0$, and so there are no bound systems from this scale and higher.

Yet, we can not  go to larger distances by our method because then $H_0^2r^2\sim 1$ and our approximation breaks down, or at best becomes questionable. In any case, there is another reason that prevents us from going to distances much larger than order $1000\ Mpc$ and it is the presence of a cosmological horizon. Since observations favour $\Omega_0\approx 1$ one sees that our metric contains a cosmological horizon at $r\approx c/H_0\approx 4400\ Mpc$; this horizon is a combination of particle horizon and cosmological event horizon and it is similar to the horizon of  static de Sitter spacetime (see for example the references  \cite{Hawking:1973,Harrison:1991} for treating horizons in cosmology). Note that the latter value of the position of the horizon is only approximate, due to our approximation, whereas the real value is larger.

\begin{figure}[]
\begin{center}
\includegraphics[bb=100 250 500 620 ,scale=0.52]{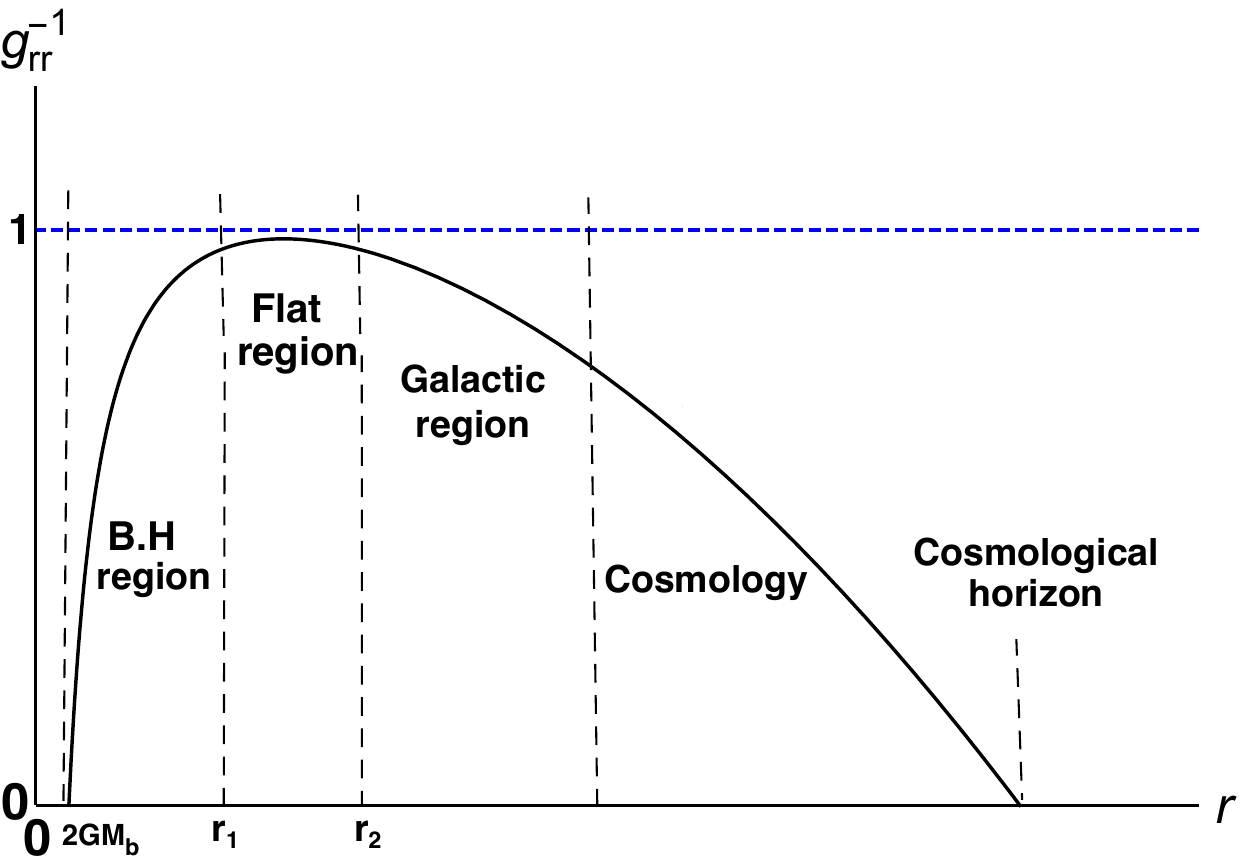}
\end{center}
\caption{\small The inverse component $g_{rr}^{-1}$ is plotted (solid line) in a schematic way (not to scale) against the radial coordinate $r$. At the black hole radius $2GM_b$ the function vanishes. The region between $r_1$ and $r_2$ is the location of the locally inertial observer: it is a flat region (with $g_{00}\approx -1$ and  $g_{rr}\approx 1$), it is far away from the black hole ($r_1>>2GM_b$), and it is small enough that the influence of the rest of the universe is negligible. In the regions $r<r_1$ and $r>r_2$ the function $g_{rr}^{-1}$ does not approximate to unity because the effects of curvature from the black hole and from the rest of the universe become important. The function $g_{rr}^{-1}$ vanishes as well at the position of the cosmological horizon.}\label{fig:inertial}
\end{figure}

A schematic plot of the function $g_{rr}^{-1}(r)$ is given in $Fig$.\ref{fig:inertial}  for the case $\Omega_0>0$. A similar schematic plot can be obtained for the redshift function $f(r)=-g_{00}(r)$ as can be checked if you take $q_0<0$.

We have chosen to plot the function $g_{rr}^{-1}$ because its zeros are the locations of  horizons in a static spacetime. In fact, we do not have to worry much about the regularity of the cosmological horizon at this stage because our approximation method is supposed to break down there and higher order terms in the derivative expansion must be taken into account.

\subsection{The Energy-Momentum Tensor}
The energy-momentum tensor can be obtained easily by inserting the above metric, $Eq$.(\ref{Fullmetric}), in the Einstein equations,
\beq
R_{\mu\nu}-\frac{1}{2}Rg_{\mu\nu}=8\pi G T_{\mu\nu}
 \eeq
 and expanding up to second order in derivatives, upon which one gets that,
\beq
T_\mu^\nu=\textrm{diagonal}(-\rho,P_r,P_\perp,P_\perp)
\eeq
where 
\beq
\rho=\frac{a_0\left(1-GM_b/r\right)}{2\pi G r}+\frac{3\Omega_0 H_0^2}{8\pi G}\left(1-\frac{2GM_b}{r}\right)\left(1-\frac{2GM_b}{3r}\right)
\eeq
\beq
P_r=-\frac{a_0M_b}{2\pi r^2}-\frac{1}{8\pi G}\left[\Omega_0H_0^2+8a_0^2-2q_0H_0^2\left(1-\frac{GM_b}{r}\right)\right]\left(1-\frac{2GM_b}{r}\right)
\eeq
\beq
P_\perp=\frac{a_0M_b}{4\pi r^2}-\frac{1}{8\pi G}\left[\Omega_0 H_0^2\left(1-\frac{GM_b}{r}\right)^2+6a_0^2-2q_0H_0^2\left(1-\frac{GM_b}{r}-\frac{G^2M^2_b}{2r^2}\right)\right]
\eeq
where in each of the three expressions, the first term is the first order term and the second term is the second order term. The first order terms were already obtained in \cite{Haddad:2020koo} and summarised in the previous section, while the second order terms are new results of this work. This energy-momentum tensor encodes information about dark matter (terms coupled to $a_0$) and dark energy (terms coupled to $H_0$, $q_0$, and $\Omega_0$) from small distances to very large distances through a nontrivial $r$ dependence.

Note that the energy-momentum tensor is anisotropic ($P_\perp\neq P_r$). However, at the largest scale (the cosmological scale) it becomes isotropic as expected. To see this let us take the large radius limit ($r\rightarrow \infty$) and obtain,

\beq
\rho=\frac{3\Omega_0H_0^2}{8\pi G}
\eeq
\beq
P_r=-\frac{1}{8\pi G}\left[\left(\Omega_0-2q_0\right)H_0^2+8a_0^2\right]
\eeq
\beq
P_\perp=-\frac{1}{8\pi G}\left[\left(\Omega_0-2q_0\right)H_0^2+6a_0^2\right]
\eeq
where an anisotropy resulting from the terms $a_0^2$ still appears. Yet, this is a small anisotropy that can be neglected compared to the $H_0^2$ terms because 
\beq
\frac{a_0^2}{H_0^2c^2}\sim 10^{-4} 
\eeq
where we have returned $c$ by unit analysis and where we have used the values $H_0=2.2\times 10^{-18}\ s^{-1}$ and $a_0\sim 10^{-11} \ m/s^2$. Therefore, at the cosmological scale we have, indeed, an isotropic energy-momentum tensor with $P_r=P_\perp\equiv P$, given by 
\beq
\rho=\frac{3\Omega_0H_0^2}{8\pi G}\qquad \qquad P=\frac{-\left(\Omega_0-2q_0\right)H_0^2}{8\pi G}
\eeq
One can check that these two last equations are simply the two Friedmann equations of cosmology, written up to second order in derivatives.
\section {The Derivative Expansion of the FLRW Metric and the Transformation to the Local Frame}
\label{RW}
The Friedmann–Lemaître–Robertson–Walker  metric is given by,
\beq\label{rw}
ds^2=-d\tau^2+R(\tau)^2\left[\frac{dr'^2}{1-kr'^2}+r'^2d\Omega^2\right]
\eeq  
where $\tau$ is the cosmological time, $R(\tau)$ is the scale factor with dimension of length, $r'$ is the dimensionless radial co-moving coordinate, $k=\{0,-1,+1\}$ is the curvature parameter, and $d\Omega^2=d\theta^2+\sin^2\theta d\phi^2$.  

In order to move to the local coordinates, in which the FLRW spacetime takes the Schwarzschild form,
\beq\label{}
ds^2=-A(t,r)dt^2+B(t,r)dr^2+r^2d\Omega^2
\eeq  
we have clearly to perform the coordinate transformation
\beq\label{}
 r=R(\tau)r'
\eeq  
to take care of the angular part, upon which the metric becomes,
\beq\label{flrw}
ds^2=-\left[1-\frac{H^2r^2}{1-kr^2/R^2}\right]d\tau^2-\frac{2Hr}{1-kr^2/R^2}d\tau dr+\frac{dr^2}{1-kr^2/R^2}+r^2d\Omega^2
\eeq 
where $H(\tau)\equiv\dot R(\tau)/R(\tau)$ is the Hubble parameter. Before we go on and cancel the cross term we find it a suitable place to stop and define the derivative expansion of the FLRW metric.

\subsection{Defining the Derivative Expansion of the FLRW Metric}
To motivate the derivative expansion method for the FLRW metric notice that significant changes of the cosmological metric occurs on large time scales (from the perspective of a local observer) and notice that significant changes in the radial direction also require large distances. In details, let us Taylor expand the scale factor $R(\tau)$ (as it encodes the time dependence of the metric) around the present time $\tau_0$,
 \beq\label{}
R(\tau)=R(\tau_0)+\dot R(\tau_0)(\tau-\tau_0)+\frac{1}{2}\ddot R(\tau_0)(\tau-\tau_0)^2+...
\eeq
or equivalently,
 \beq\label{}
R(\tau)=R_0\left[1+H_0(\tau-\tau_0)-\frac{1}{2}q_0H_0^2(\tau-\tau_0)^2+...\right]
\eeq
where as usual $R_0\equiv R(\tau_0)$ is the scale factor at the present time, $H_0\equiv \dot{R_0}/R_0$ is the Hubble's constant, and $q_0\equiv -\ddot R_0 R_0/\dot R_0^2$ is the deceleration parameter. Pay attention that since $H_0$ is very small the hierarchy in the Taylor terms
 \beq\label{}
 R(\tau_0)>>\dot R(\tau_0)(\tau-\tau_0)>>\frac{1}{2}\ddot R(\tau_0)(\tau-\tau_0)^2
\eeq
continues to hold even for large time intervals $\tau-\tau_0$ and that is due to the hierarchy in the derivatives
 \beq\label{}
 O(1)>>O(\partial_\tau)=O(H_0)>>O(\partial^2_\tau)=O(H_0^2)
\eeq

As for the radial direction notice that the coordinate $r$ appears in the metric, $Eq$.(\ref{flrw}), only through the ratio $r/R$,\footnote{For instance $Hr=\dot Rr/R$.} where we are excluding the angular part $r^2d\Omega^2$ from our discussion as it is already in the Schwarzschild form and so we will leave it as it is. Since the scale factor $R$ is very large\footnote{We are talking about periods in the universe when the scale factor is very large. Our analysis breaks down if the scale factor is small enough.}  it is clear that significant changes along the radial direction $r$ require large distances for a local observer. More precisely, if we Taylor expand any component of the metric $g(r)$ around $r=0$ we get,
 \beq\label{}
g(r)=g(0)+\partial_r g(0)r+ \frac{1}{2}\partial^2_r g(0) r^2+...=g(0)+\frac{\partial_x g(0)}{R}r+ \frac{1}{2}\frac{\partial^2_x g(0)}{R^2}r^2+...
\eeq
where $x=r/R$. Thus, similar to the case of time derivatives discussed above, here as well the generic hierarchy in the Taylor terms 
 \beq\label{}
 g(0)>>\partial_r g(0)r>>\frac{1}{2}\partial^2_r g(0)r^2
\eeq
continues to hold even for large $r$'s and that is because of the hierarchy in the derivatives
 \beq\label{}
 O(1)>>O(\partial_r)=O(1/R)>>O(\partial^2_r)=O(1/R^2)
\eeq

One more comment deserves being made. By looking at one of the Friedman equations,
 \beq\label{}
 \frac{\dot R^2}{R^2}=\frac{8\pi G}{3}\rho-\frac{k}{R^2}
\eeq
one can see that generically the term $k/R^2$ is of the same order as $\dot R^2/R^2$ because they are terms in the same equation. Therefore, since $k/R^2=O(\partial^2_r)$ and $\dot R^2/R^2=H^2=O(\partial^2_\tau)$ we conclude that the time derivative and the radial derivative are generically of the same order,
  \beq\label{}
O(\partial_r)=O(\partial_\tau)
\eeq

\subsection{The FLRW Spacetime in the Local Frame}
Now we can go on and expand the FLRW metric Eq.(\ref{flrw}) up to second order in derivatives and obtain,
\beq\label{cross}
ds^2=-\left[1-H_0^2r^2\right]d\tau^2-2\left[H_0+\dot H_0(\tau-\tau_0)\right]r d\tau dr+\left[1+kr^2/R_0^2\right]dr^2+r^2d\Omega^2
\eeq 
To cancel the cross term we perform a general coordinate transformation of the form 
\beq\label{}
\tau=\tau(t,r)
\eeq 
after which we get (all subsequent calculations are going to be up to second order in derivatives),
\begin{align}\label{long}\nonumber
ds^2&=-\left[1-H_0^2r^2\right]\left(\frac{\partial\tau}{\partial t}\right)^2dt^2-2\frac{\partial\tau}{\partial t}\left(\left[H_0+\dot H_0(\tau-\tau_0)\right]r+\left[1-H_0^2r^2\right]\frac{\partial \tau}{\partial r}\right)dtdr\\ 
&+\left(1+kr^2/R_0^2-\left[1-H_0^2r^2\right]\left(\frac{\partial\tau}{\partial r}\right)^2-2\left[H_0+\dot H_0(\tau-\tau_0)\right]r\frac{\partial\tau}{\partial r}\right)dr^2+r^2d\Omega^2
\end{align}
To achieve $g_{tr}=0$ we must require that,
\beq\label{}
\frac{\partial\tau}{\partial r}=-\left[H_0+\dot H_0(\tau-\tau_0)\right]r
\eeq 
Upon integration with respect to $r$ this yields, 
\beq\label{co-tr}
\tau=-\frac{1}{2}\left[H_0+\dot H_0(\tau-\tau_0)\right]r^2+\psi(t)
\eeq 
where $\psi(t)$ is an arbitrary (gauge) function of $t$. By inserting equation Eq.(\ref{co-tr}) into Eq.(\ref{long}) one obtains,
 \beq\label{short}
ds^2=-\left[\left(\frac{d\psi}{dt}\right)^2-\left(H_0^2\left(\frac{d\psi}{dt}\right)^2+\dot H_0 \frac{d\psi}{dt}\right)r^2\right]dt^2+\left[1+\left(H_0^2+\frac{k}{R_0^2}\right)r^2\right]dr^2+r^2d\Omega^2
\eeq
We will choose the gauge,
 \beq\label{}
d\psi/dt=1
\eeq
 in order to make $g_{tt}=-1$ at very small $r$'s since we are talking about the coordinates of a locally inertial observer. One can also easily check that,
 \beq\label{}
\dot H_0=-\left(1+q_0\right)H_0^2
\eeq
 All in all, the metric Eq.(\ref{short}) reads now,
 \beq\label{}
ds^2=-\left(1+q_0H_0^2r^2\right)dt^2+\frac{dr^2}{1-\left(H_0^2+\frac{k}{R_0^2}\right)r^2}+r^2d\Omega^2
\eeq
which has the required form. The last step is going to be for the sake of appearance only,  and it is to use the definition of the dimensionless density parameter $\Omega_0=\left(H_0^2+\frac{k}{R_0^2}\right)/H_0^2$ and to rewrite the above metric in the final desired form,
\beq\label{}
ds^2=-\left(1+q_0H_0^2r^2\right)dt^2+\frac{dr^2}{1-\Omega_0H_0^2r^2}+r^2d\Omega^2
\eeq

This is the FLRW spacetime written in the local coordinates $(t,r)$ up to second order in derivatives. It is to be said that this metric was found before in \cite{Mizony:2004sh} by using a different approximation scheme. It is important to notice that the metric is static; the time-dependence does not appear at this order. The above FLRW metric (with a precision of order $H_0^2$) is supposed to cover many observations in cosmology. As will be shown in the next section this metric covers the region of small redshifts with $z<1$. For large redshifts, that is, for $z>1$, we must restore to the full FLRW metric.
\section {The Cosmological Redshift from the Local Metric}
\label{redsh}
In this section we firstly show how the cosmological redshift $z$ is obtained from the local metric and that it coincides with the familiar formula in the literature up to order $H_0^2$. Secondly, we show that up to this order the cosmological redshift is the sum of two effects: a gravitational shift and a special-relativistic Doppler redshift. 

\subsection {Calculation of the Cosmological Redshift}
At cosmological distances the baryonic contributions become negligible and the local metric $Eq$.(\ref{Fullmetric}) reads,
\beq\label{}
ds^2=-fdt^2+hdr^2+r^2d\Omega^2=-\left(1+2a_0r+q_0H_0^2r^2\right)dt^2+\frac{dr^2}{1-2a_0r-\Omega_0 H_0^2r^2}+r^2d\Omega^2
\eeq  
where we have neglected all terms containing the baryonic mass $M_b$ and kept the terms of dark matter and dark energy.

Take a galaxy with 4-velocity $V^\mu$ which sends a light signal toward us (relative to such large distances we are almost at the centre of the local coordinates $r=0$) with a wave vector $K^\mu=(K^0,K^r,0,0)$. The frequency of the emitted light as measured by the galaxy (the source) is $\omega_s=-K_\mu V^\mu$ evaluated at the position and time of emission $(t,r,\theta,\phi)$. Similarly, the frequency observed on Earth is $\omega_o=-K_\mu U^\mu$ evaluated at the position and time of observation $(t=0,r=0)$, where $U^\mu=(1,\vec{0})$ is the 4-velocity of the stationary observer situated at $r=0$. After using the conditions $V_\mu V^\mu=-1$ and $K_\mu K^\mu=0$ one obtains for the ingoing null ray and for the galaxy respectively,
\beq\label{}
 K^r=-\sqrt{\frac{f}{h}}K^0 \qquad \textrm{and} \qquad  V^0=\frac{1}{\sqrt{f-h(\frac{dr}{dt})^2-r^2(\frac{d\phi}{dt})^2}}
\eeq
where we have used the relations $V^0=\frac{dt}{d\tau}$, $V^r=\frac{dr}{d\tau}=\frac{dt}{d\tau}\frac{dr}{dt}=V^0\frac{dr}{dt}$,  $V^\phi=\frac{d\phi}{d\tau}=V^0\frac{d\phi}{dt}$, and $V^\theta=0$ \footnote{The path of the galaxy takes place on a single plane, $\theta=0$.}$^{,}$\footnote{In this section $\tau$ is the proper time.}. Since our metric is static with a corresponding killing vector $\xi^\mu=(dt)^\mu$ the energy of particles is conserved and we have $K_\mu\xi^\mu=\textrm{constant}$; this gives that $K^0(r)=K^0(0)/f(r)$.

Thus the frequency of the light ray as measured by the galaxy is,
\beq\label{}
\omega_s=-K_\mu V^\mu |_{(t,r)}=\frac{K^0(0)}{f(r)}\frac{f(r)+\sqrt{f(r)h(r)} \frac{dr}{dt}}{\sqrt{f(r)-h(r)(\frac{dr}{dt})^2-r^2(\frac{d\phi}{dt})^2}}
\eeq  
and the observed frequency on Earth is,
\beq\label{}
\omega_o=-K_\mu U^\mu |_{(t=0,r=0)}=K^0(0)
\eeq  
The redshift is therefore,
\beq
z=\frac{\omega_s}{\omega_o}-1=\frac{1}{\sqrt{f}}\frac{1+\sqrt{\frac{h}{f}} \frac{dr}{dt}}{\sqrt{1-\frac{h}{f}(\frac{dr}{dt})^2-\frac{r^2}{f}(\frac{d\phi}{dt})^2}}-1
\eeq  
By writing $dr/dt=v_r$ and $rd\phi/dt=v_\phi$ the redshift formula becomes,
\beq\label{red}
z=\frac{1}{\sqrt{f}}\frac{1+\sqrt{\frac{h}{f}}v_r}{\sqrt{1-\frac{h}{f}v_r^2-\frac{1}{f}v_\phi^2}}-1
\eeq  
If the spacetime were flat, $f=h=1$, we would recover the special relativistic Doppler effect,
\beq\label{}
z=\frac{1+ v_r}{\sqrt{1-(\vec{v})^2}}-1
\eeq  
If the emitting source were static we would obtain the purely gravitational shift, 
\beq\label{}
z=\frac{1}{\sqrt{f}}-1
\eeq

Going back to the full formula, $Eq.$(\ref{red}), recall that in this particle-dynamics treatment of galaxies the initial distances and initial velocities of  galaxies must be supplied by us as part of the initial value problem. Notice that the $r$, $v_r$, and $v_\phi$, appearing in the redshift formula $Eq.$(\ref{red}) are the coordinate distance of the galaxy and its velocity components at the time of emission and hence can be taken as initial values. Galaxies in clusters are commonly known "not" to have rotation about the centre of the cluster, that is, $v_\phi\approx 0$, while the radial velocity is observed to correspond to the Hubble flow plus some peculiar velocity\footnote{Note that the expansion of the universe $v_r=H_0r$ is not derived (or predicted) by our analysis but must be given in advance as part of the initial values set.},
\beq\label{initialv}
 v_r=H_0r+v_{pec}
 \eeq 
 It is to be said before we proceed that in this section the use of the coordinate distance $r$ or the proper distance $D$ to the galaxy (at the time of emission) does not change the results, because as can be checked,
 \beq\label{}
 D=\int_0^r\sqrt{h(r)}dr=r+O(a_0)
 \eeq 
 which would give a correction of order $O(a_0H_0)$ to the Hubble's law which is smaller than the order $O(H_0^2)$, up to which we want to work; one can check that $a_0/H_0\sim 10^{-2}$.
 
 Now if we expand the full expression of the redshift $Eq$.(\ref{red}) up to second order in derivatives by using our initial conditions  $v_\phi\approx 0$ and $v_r=H_0r+v_{pec}$ (we are interested in cases where the peculiar velocity is very small compared to the Hubble flow, $v_{pec}=O(H_0^2r^2)$ or smaller), then we obtain after some work,
\begin{align}\label{}
z=H_0\left(1-\frac{a_0}{H_0}\right)r+\frac{1}{2}\left(1-q_0-2\frac{a_0}{H_0}+3\frac{a_0^2}{H_0^2}\right)H_0^2r^2+v_{pec}
\end{align}
In the second parenthesis $a_0^2/H_0^2\sim 10^{-4}$  and $a_0/H_0\sim 10^{-2}$ and so can be neglected since they give rise to terms much smaller than order $H_0^ 2$, and so,

\begin{align}\label{}
z=H_0\left(1-\frac{a_0}{H_0}\right)r+\frac{1}{2}\left(1-q_0\right)H_0^2r^2+v_{pec}
\end{align}
The ratio $a_0/H_0\sim 10^{-2}$ is within the error in determining the value of Hubble's constant, but nevertheless, it can not be ignored, in principle, if we are going to next order terms. Instead, we can rescale the Hubble's constant by defining the observed Hubble's constant,
\beq\label{}
H_0^{obs}\equiv H_0\left(1-\frac{a_0}{H_0}\right)
\eeq  
after which one obtains,
\begin{align}\label{redshift}
z=H^{obs}_0r+\frac{1}{2}\left(1-q_0\right)(H_0^{obs})^2r^2+v_{pec}
\end{align}
Since, as stated above, the initial peculiar velocity $v_{pec}$ is of the order of $H_0^2r^2$ or smaller we can write it as
\beq\label{}
 v_{pec}=\frac{C}{2}(H_0^{obs})^2r^2
 \eeq  
 where $C$ is a constant that must be determined by the initial conditions. Upon this we obtain,
 \begin{align}\label{redshift}
z=H^{obs}_0r+\frac{1}{2}\left(1-q_0+C\right)(H_0^{obs})^2r^2
\end{align}
 If this last equation is inverted one obtains $r$ in terms of $z$,
\beq\label{}
r=\frac{1}{H_0^{obs}}\left[z-\frac{1}{2}\left(1-q_0+C\right)z^2\right]
\eeq  
To coincide with the literature (see for example \cite{Weinberg:1972kfs,Weinberg:2008zzc,Ryden:2003yy}) we must fix 
\beq\label{}
C=2q_0
 \eeq  
 upon which one gets the familiar distance-redshift relation for small redshifts ($z<1$),
 
 \beq\label{}
r=\frac{1}{H_0^{obs}}\left[z-\frac{1}{2}\left(1+q_0\right)z^2\right]
\eeq  
which is the goal of this section.

It is worth to conclude this section with the following comment. After fixing the constant $C$ the velocity of the source galaxy at the time of emission, $Eq$.(\ref{initialv}), can be written as
\beq\label{}
 v_r=H_0r+q_0H_0^2r^2
 \eeq 
 where it does not differ whether we used $H_0$ or $H_0^{obs}$ as long as we are working up to order $H_0^2$. Since the light was sent from the source galaxy at negative time $t$ and received on Earth at time $t=0$ one can write $r=-t$ for the distance to the galaxy. Then one can write $v_r=v_{0}+at$, describing motion at constant acceleration in flat spacetime, where $v_0=H_0r$ is the initial velocity and $a=-q_0H_0^2r$ is the cosmological acceleration -- see $Eq$.(\ref{cos-acc}). This comment shows that the value we have fixed for the constant $C$ is consistent with the physical cosmological picture as it tells that the galaxy is accelerating at the rate of the cosmological acceleration of the universe.
\subsection {The Cosmological Redshift is the Sum of Doppler and Gravitational Shifts}
The special-relativistic Doppler shift from a source receding from a static observer at a constant velocity along the line of sight is given by,
\beq\label{}
z_{dop}=\frac{\omega_s}{\omega_o}-1=\sqrt{\frac{1+\beta}{1-\beta}}-1=\beta +\frac{1}{2}\beta^2+...
\eeq
where $\beta$ is the velocity of the source and where we have expanded up to order $\beta^2$. In our case $\beta$ is the recessional velocity of some galaxy (that is, $\beta=H_0r+v_{pec}$) and so,
\beq\label{}
z_{dop}=H_0r +\frac{1}{2}H_0^2r^2+v_{pec}+...
\eeq
where we have assumed as usual $v_{pec}<<H_0r$.
The gravitational shift, on the other hand, in a light ray sent from a static source at distance $r$ and received by an observer at $r=0$ in our static spacetime is given by,
\beq\label{}
z_{grav}=\frac{1}{\sqrt{f(r)}}-1=-a_0r+\frac{1}{2}\left(-q_0+3\frac{a_0^2}{H_0^2}\right)H_0^2r^2+...
\eeq
The total frequency shift is the sum of the above two shifts,
\beq\label{}
z=z_{dop}+z_{grav}=H_0\left(1-\frac{a_0}{H_0}\right)r+\frac{1}{2}\left(1-q_0+3\frac{a_0^2}{H_0^2}\right)H_0^2r^2+v_{pec}
\eeq
By repeating the same approximations, the same rescaling of the Hubble's constant, and the same treatment for $v_{pec}$, done in the previous section, we reproduce exactly the same cosmological redshift $z$ given in Eq.(\ref{redshift}) which is the goal of this section. The redshift analysis of the previous section is more general and more complete than the one performed here. However, the different effects contributing to the redshift are entangled in one formula, Eq.(\ref{red}), and are not split neatly from each other. The goal of this section was to split the contributing effects from each other and to identify each of them separately. A similar analysis was done for example in \cite{Whiting:2004ds} but was restricted to specific cosmologies;  the authors stated that the breaking of the full effect into two components (doppler and static gravitational shift) can be obtained in static universes only. Here we have shown that this breaking occurs for all cosmologies as long as we are working up to order $H_0^2$.

\section {Discussion} 
\label{Diss}

The main result of this paper is to put the three significant scales - the solar system, the galactic, and the cosmological scales - in one framework. The unifying framework is argued to be the derivative expansion method. This framework can be seen as the natural one if we think from the point of view of a locally inertial observer, like us, as was discussed in detail in the bulk of the paper. In what follows we are going to give several comments that are thought to be important:
\begin{enumerate}

\item The galactic scale was missed in the previous works trying to connect the solar system with  cosmology -- see for instance \cite{McVittie:1933zz,Einstein:1945id,Cooperstock:1998ny,Mizony:2004sh,Faraoni:2007es,Mitra:2013rma}. In this work it is shown that the galactic scale (first order in derivatives) connects the latter two scales. It is to be said that linear terms (in $r$) in the metric components $g_{00}$ and $g_{rr}$ were never correlated before \cite{Haddad:2020koo} with dark matter, in contrast to quadratic terms (proportional to $r^2$) which were long ago correlated with dark energy. The consistency of the approach together with its success at giving the main observations in each scale give further support that those linear terms are indeed related to dark matter.  

\item This framework suggests a new possible view of dark matter and dark energy. Dark matter and dark energy could be different "projections" of the same matter field. That is, there could be a single dark field in space that changes very slowly, and so when it is Taylor expanded -- that is, when we expand  $V(r)$ and $U(r)$ -- the first order terms are what we call dark matter and the second order terms are what we call dark energy. 

\item When we talk about some distance $r$ then by the parameter $M_b$ we mean the baryonic mass inside this radius. If $r$ is just outside the solar system then $M_b$ is almost the mass of the Sun, if $r$ is just outside a galaxy then $M_b$ is the baryonic mass of the galaxy, and if $r$ is just outside a cluster of galaxies then $M_b$ is the baryonic mass of the cluster. It is interesting to note how our method was applied also to galaxy clusters. In $Sec.[$\ref{Overview}] it was shown that our analysis predicts that galaxies in clusters are freely falling toward the centre of the cluster with a constant acceleration $a_0\sim 10^{-11}\ m/s^2$, and at the same time moving collectively along with the Hubble's flow.

\item The spacetime proposed in this work (up to second order in derivatives)  is anisotropic and the anisotropy is expressed by the result that the radial and perpendicular pressures of dark matter and dark energy are unequal, $P_r\neq P_\perp$, in general. It is interesting to see how our metric restores the isotropy when we take the limit of very large $r$. The same comment holds also for the energy-momentum tensor. 

\item Finally, we would like to stress that our constructed metric is a local one (corresponding to small redshifts, $z<1$). Yet our local metric is valid up to great distances, up to distances that are only one order of magnitude smaller than the distance to the cosmological horizon (see $Sec.$ \ref{Overview} and the discussion about validity).
In numbers, our local metric is valid up to distances of order $1000\ Mpc$ (corresponding to $z<1$) while the distance to the cosmological horizon is of order $10,000 \ Mpc$. Distances corresponding to large $z$ are packed close to the cosmological horizon.  
 
In order to discuss large redshifts ($z>1$) the time-dependence of the FLRW metric must be taken into account since such redshifts correspond to light rays sent at times when the scale factor of the universe was considerably smaller than at present. Thus our static local metric is not valid in this regime. In this case one simply has to turn to the FLRW metric which is our boundary condition. However, if one wishes one can continue with the derivative expansion method up to third order and higher but allow now for the time-dependence to enter; that is, one has to perform the derivative expansion in both the $r$ and $t$ coordinates and one has to match the local metric with the global FLRW metric in the matching region. From the local metric point of view the matching region is the large radius limit. The matching is done by fixing the higher order coefficients of the local metric so as to coincide with the FLRW metric (expanded up to the relevant order).

\end{enumerate}

\appendix
\section{Eddington-Finkelstein Coordinates and Regularity of the Horizon}
\label{app:Edd}

Since we have spherical symmetry, in order to find the location of the horizon we look for the null $r=\textrm{constant}$ surface,
\beq\label{}
g^{\mu\nu}\partial_\mu r \partial_\nu r=0
\eeq 
which for our metric $Eq$.(\ref{metric1}) gives

\beq\label{}
g^{rr}=h^{-1}=0
\eeq 
To make the horizon regularity manifest we move to the new coordinate $v=t+r_*$ where $dr_*/dr=\sqrt{h/f}$, upon which our metric $Eq$.(\ref{metric1}) becomes
\beq\label{}
ds^2=-f(r)dv^2+2\sqrt{f(r)h(r)}dvdr+r^2\left(d\theta^2+\sin^2\theta d\phi^2\right)
\eeq 
In order to prevent a singularity in $g_{rv}$ at the horizon (as $h\rightarrow \infty$) clearly we must have
 
 \beq\label{}
f(r)=h(r)^{-1} \qquad \textrm{as}\qquad r\rightarrow r_0
\eeq

\section{Geodesic Equation in Spherically Symmetric Static Spacetimes}
\label{app:geod}
For the general spherically symmetric static metric 
\beq\label{}
ds^2=-f(r)dt^2+h(r)dr^2+r^2\left(d\theta^2+\sin^2\theta d\phi^2\right)
\eeq  
the geodesic equations are 
\beq\label{}
\frac{d^2x^\mu}{dp^2}+\Gamma^{\mu}_{\nu\lambda}\frac{dx^\nu}{dp}\frac{dx^\lambda}{dp}=0
\eeq  
where $p$ is a parameter along the trajectory. Because of spherical symmetry the motion will take place in a plane, and without loss of generality we will take it to be in the $\theta=\pi/2$ plane.
The radial equation of motion will be (here we are going to follow reference \cite{Weinberg:1972kfs})
 \beq\label{}
\frac{d^2r}{dp^2}+\frac{h'}{2h}\left(\frac{dr}{dp}\right)^2-\frac{J^2}{r^3h}+\frac{f'}{2hf^2}=0
\eeq 
 where the prime denotes derivative with respect to $r$. There are also the two familiar relations
 
 \beq\label{geod2}
\frac{dt}{dp}=\frac{1}{f(r)} \qquad \textrm{and} \qquad r^2\frac{d\phi}{dp}=J
\eeq 
where the constant $J$ is the angular momentum per unit mass.  In terms of the time coordinate $t$ the radial equation becomes,
\beq\label{geod1}
\frac{d^2r}{dt^2}+\left(\frac{h'}{2h}-\frac{f'}{f}\right)\left(\frac{dr}{dt}\right)^2-\frac{f^2J^2}{r^3h}+\frac{f'}{2h}=0
\eeq

 For a circular motion, $r=\textrm{constant}$, the equation \ref{geod1} simply becomes
 \beq\label{geod3}
\frac{J^2}{r^3}=\frac{f'}{2f^2}
\eeq 
On the other hand, upon combining the two equations in \ref{geod2} one gets that 
 \beq\label{geod4}
J=r^2\frac{d\phi}{dt}f^{-1}
\eeq 
Finally, if we insert equation \ref{geod4} into equation \ref{geod3} we get
 \beq\label{geod5}
\frac{(v_\phi)^2}{r}=\frac{1}{2}f'
\eeq 
where $v_\phi=rd\phi/dt$ is the angular velocity. Thus, finally, the rotation curve for circular orbits takes the simple form,
 \beq\label{geod5}
(v_\phi)^2=\frac{1}{2}f'r
\eeq


\begin{thebibliography}{99}

\bibitem{McVittie:1933zz}
G.~C.~McVittie,
``The mass-particle in an expanding universe,''
Mon. Not. Roy. Astron. Soc. \textbf{93} (1933), 325-339
doi:10.1093/mnras/93.5.325


\bibitem{Einstein:1945id}
A.~Einstein and E.~G.~Straus,
``The influence of the expansion of space on the gravitation fields surrounding the individual stars,''
Rev. Mod. Phys. \textbf{17} (1945), 120-124
doi:10.1103/RevModPhys.17.120



\bibitem{Cooperstock:1998ny}
F.~I.~Cooperstock, V.~Faraoni and D.~N.~Vollick,
``The Influence of the cosmological expansion on local systems,''
Astrophys. J. \textbf{503} (1998), 61
doi:10.1086/305956
[arXiv:astro-ph/9803097 [astro-ph]].



\bibitem{Mizony:2004sh}
M.~Mizony and M.~Lachieze-Rey,
``Cosmological effects in the local static frame,''
Astron. Astrophys. \textbf{434} (2005), 45-52
doi:10.1051/0004-6361:20042195
[arXiv:gr-qc/0412084 [gr-qc]].


\bibitem{Faraoni:2007es}
V.~Faraoni and A.~Jacques,
``Cosmological expansion and local physics,''
Phys. Rev. D \textbf{76} (2007), 063510
doi:10.1103/PhysRevD.76.063510
[arXiv:0707.1350 [gr-qc]].

\bibitem{Mitra:2013rma}
A.~Mitra,
``Friedmann-Robertson-Walker metric in curvature coordinates and its applications,''
Grav. Cosmol. \textbf{19} (2013), 134-137
doi:10.1134/S0202289313020072


\bibitem{Manasse:1963zz}
F.~K.~Manasse and C.~W.~Misner,
``Fermi Normal Coordinates and Some Basic Concepts in Differential Geometry,''
J. Math. Phys. \textbf{4} (1963), 735-745
doi:10.1063/1.1724316


\bibitem{Ni:1978zz}
W.~T.~Ni and M.~Zimmermann,
``Inertial and gravitational effects in the proper reference frame of an accelerated, rotating observer,''
Phys. Rev. D \textbf{17} (1978), 1473-1476
doi:10.1103/PhysRevD.17.1473

\bibitem{Marzlin:1994wc}
K.~P.~Marzlin,
``On the physical meaning of Fermi coordinates,''
Gen. Rel. Grav. \textbf{26} (1994), 619
doi:10.1007/BF02108003
[arXiv:gr-qc/9402010 [gr-qc]].

\bibitem{Misner:1973prb}
C.~W.~Misner, K.~S.~Thorne and J.~A.~Wheeler,
``Gravitation,''



\bibitem{Pajer:2013ana}
E.~Pajer, F.~Schmidt and M.~Zaldarriaga,
``The Observed Squeezed Limit of Cosmological Three-Point Functions,''
Phys. Rev. D \textbf{88} (2013) no.8, 083502
doi:10.1103/PhysRevD.88.083502
[arXiv:1305.0824 [astro-ph.CO]].


\bibitem{Dai:2015jaa}
L.~Dai, E.~Pajer and F.~Schmidt,
``On Separate Universes,''
JCAP \textbf{10} (2015), 059
doi:10.1088/1475-7516/2015/10/059
[arXiv:1504.00351 [astro-ph.CO]].


\bibitem{Cabass:2016cgp}
G.~Cabass, E.~Pajer and F.~Schmidt,
``How Gaussian can our Universe be?,''
JCAP \textbf{01} (2017), 003
doi:10.1088/1475-7516/2017/01/003
[arXiv:1612.00033 [hep-th]].

\bibitem{Dai:2015rda}
L.~Dai, E.~Pajer and F.~Schmidt,
``Conformal Fermi Coordinates,''
JCAP \textbf{11} (2015), 043
doi:10.1088/1475-7516/2015/11/043
[arXiv:1502.02011 [gr-qc]].


\bibitem{Tully:1977fu}
  R.~B.~Tully and J.~R.~Fisher,
  ``A New method of determining distances to galaxies,''
  Astron.\ Astrophys.\  {\bf 54} (1977) 661.


\bibitem{McGaugh:2000sr}
  S.~S.~McGaugh, J.~M.~Schombert, G.~D.~Bothun and W.~J.~G.~de Blok,
  ``The Baryonic Tully-Fisher relation,''
  Astrophys.\ J.\  {\bf 533} (2000) L99
  doi:10.1086/312628
  [astro-ph/0003001].


 
\bibitem{McGaugh:2011ac}
  S.~McGaugh,
  ``The Baryonic Tully-Fisher Relation of Gas Rich Galaxies as a Test of LCDM and MOND,''
  Astron.\ J.\  {\bf 143} (2012) 40
  doi:10.1088/0004-6256/143/2/40
  [arXiv:1107.2934 [astro-ph.CO]].



\bibitem{Weinberg:1972kfs}
  S.~Weinberg,
  ``Gravitation and Cosmology : Principles and Applications of the General Theory of Relativity'', (New York : Wiley, 1972), isbn: 978-0-471-92567-5



\bibitem{Wald:1984rg}
  R.~M.~Wald,
  ``General Relativity,''
  doi:10.7208/chicago/9780226870373.001.0001

\bibitem{Haddad:2020koo}
F.~Haddad and N.~Haddad,
``A Black Hole inside Dark Matter and the Rotation Curves of Galaxies,'' International Journal of Modern Physics D $|$ Vol. 29, No. 15, 2050107 (2020),
doi:10.1142/S0218271820501072
[arXiv:2002.12772 [gr-qc]].





\bibitem{Sofue:2017}
Y.~Sufue, "Rotation and mass in the Milky Way and spiral galaxies", Publications of the Astronomical Society of Japan, Volume 69, Issue 1, February 2017, R1, https://doi.org/10.1093/pasj/psw103


\bibitem{Sofue:2015}
Y.~Sufue, "Dark halos of M31 and the Milky Way", Publications of the Astronomical Society of Japan, Volume 67, Issue 4, August 2015, 75, 
 https://doi.org/10.1093/pasj/psv042

\bibitem{Milgrom:2014usa}
M.~Milgrom,
``MOND theory,''
Can. J. Phys. \textbf{93} (2015) no.2, 107-118
doi:10.1139/cjp-2014-0211
[arXiv:1404.7661 [astro-ph.CO]].

\bibitem{Milgrom:2019cle}
M.~Milgrom,
``MOND vs. dark matter in light of historical parallels,''
Stud. Hist. Phil. Sci. B \textbf{71} (2020), 170-195
doi:10.1016/j.shpsb.2020.02.004
[arXiv:1910.04368 [astro-ph.GA]].


\bibitem{Bowers:1974tgi}
  R.~L.~Bowers and E.~P.~T.~Liang,
  ``Anisotropic Spheres in General Relativity,''
  Astrophys.\ J.\  {\bf 188} (1974) 657.
  doi:10.1086/152760

\bibitem{Weinberg:2008zzc}
S.~Weinberg,
``Cosmology,'' Published in: Oxford, UK: Oxford Univ. Pr. (2008) 593 p
ISBN: 9780198526827


\bibitem{Ryden:2003yy}
B.~Ryden,
``Introduction to cosmology,'' Published in: San Francisco, USA: Addison-Wesley (2003) 244 p
ISBN: 9781107154834 (Print), 9781316889848


\bibitem{Whiting:2004ds}
A.~B.~Whiting,
``The Expansion of space: Free particle motion and the cosmological redshift,''
Observatory \textbf{124} (2004), 174
[arXiv:astro-ph/0404095 [astro-ph]].
  
\bibitem{Hawking:1973}
  S.~Hawking, G.~Ellis, (1973). "The Large Scale Structure of Space-Time" (Cambridge Monographs on Mathematical Physics). Cambridge: Cambridge University Press. doi:10.1017/CBO9780511524646   


\bibitem{Harrison:1991}
E.~Harrison, 
"Hubble Spheres and Particle Horizons",
(1991) Astrophysical Journal v.383, p.60,
doi:10.1086/170763





  

  
  
  
  


	

     
\end{thebibliography}
\end{document}